\documentclass{iopconfser}
\usepackage{float}
\usepackage{subcaption}
\usepackage{graphicx}
\usepackage{wrapfig}

\usepackage[backend=biber,sorting=none,style=nature,autocite=superscript]{biblatex}
\begin{document}

\title{Effect of Stokes number on erosion on Pelton buckets for sediment-laden flows}

\author{Aron Dagur Beck$^{1}$ and Elena Vagnoni$^{1}$}

\affil{$^1$Technology Platform for Hydraulic Machines, EPFL, Lausanne, Switzerland}

\email{aron.beck@epfl.ch}

\begin{abstract}
With increased glacial melting and the need to maintain sediment continuity for ecosystem health, sediment-laden flows through hydropower plants are becoming increasingly problematic, particularly due to erosion on runner blades and buckets. A widely used mitigation strategy is the use of filters to protect Pelton turbines. However, these filters lead to rapid sediment accumulation in reservoirs, which must be drained frequently to maintain storage capacity. The high cost of such drainage operations calls for longer intervals between them, without compromising runner integrity, as erosion-induced stress concentrations may cause bucket rupture.
A better understanding of the causes of runner erosion under varying operational and sediment conditions is essential to allow more sediment to pass through the hydraulic machine safely. This study investigates how sediment-laden flow through the nozzle affects particle distribution in the jet for different Stokes numbers. Furthermore, it analyses how a realistic particle size and spatial distribution in the impacting jet compares to the assumption of a uniform particle distribution with particles of mean size when simulating bucket erosion in Pelton wheels.
The results show that the particle distribution in the jet follows a similar axisymmetric shape for low Stokes numbers whereas at higher Stokes numbers it becomes asymmetric. Additionally, it is shown that imposing uniform particle distribution with particles of mean diameter under-predicts tip- and splitter erosion when simulating the erosion on the bucket but is captured by imposing the realistic size and spatial distribution. 
\end{abstract}

\section{Introduction}
The effects of climate change are increasingly evident in the hydropower field, both through the more frequent transients required to balance the grid and the rise in sediment transport due to glacial melting. These sediments can lead to significant erosion of hydraulic equipment if allowed to flow freely through the turbine. At the Bitsch power plant in Switzerland, for instance, filters are currently employed to reduce sediment intake, which leads to rapid sediment build-up in the reservoir and requires frequent and costly drainage interventions. Consequently, flushing sediments through the hydraulic machine is drawing increasing attention by the plant operators, and understanding the impact of sediments on turbine components is of growing importance.
Sediment erosion in Pelton turbines has been studied both experimentally~\cite{padhy_effect_2009, rai_effect_2020} and numerically~\cite{xiao_analysis_2022, ge_experimental_2021, liu_research_2025}. In the work by Ge et al.~\cite{ge_experimental_2021}, particle-laden flow was simulated using an Eulerian-Lagrangian framework, where the air–water interface was resolved using the Volume of Fluid (VOF) method, and particle trajectories were tracked using the Discrete Phase Model (DPM). In that study, both the jet velocity and the particle spatial distribution at the inlet were assumed to be uniform.
Xiao et al.~\cite{xiao_analysis_2022} employed a similar methodology, but included the flow through the nozzle in the simulation to capture a more realistic jet velocity and particle distribution. This allowed them to identify how sediment erosion varies with the bucket rotation angle at the best efficiency point (BEP). Liu et al.~\cite{liu_research_2025} further extended the approach by analysing the impact of different particle sizes on the erosion rate of the buckets. Additionally, other studies have simulated the flow through the nozzle specifically to investigate erosion in the injector region~\cite{tarodiya_particulate_2022, guo_sediment-laden_2020}.
However, it remains unclear how the particle distribution within the jet influences erosion on the bucket, and how the key erosion-driving parameters can be scaled. To address this gap, we investigate how the particle Stokes number affects the particle distribution in the jet, with the aim of gaining deeper insight into the mechanisms and scaling laws behind bucket erosion. Using 3D numerical models, we compare the erosion patterns that result from assuming a uniform particle distribution to those obtained using a realistic distribution shaped by the flow through the nozzle.

\begin{wrapfigure}{R}{0.4\textwidth}
    \centering
    \begin{subfigure}[b]{0.4\textwidth}
        \centering
        \includegraphics[width=1\linewidth]{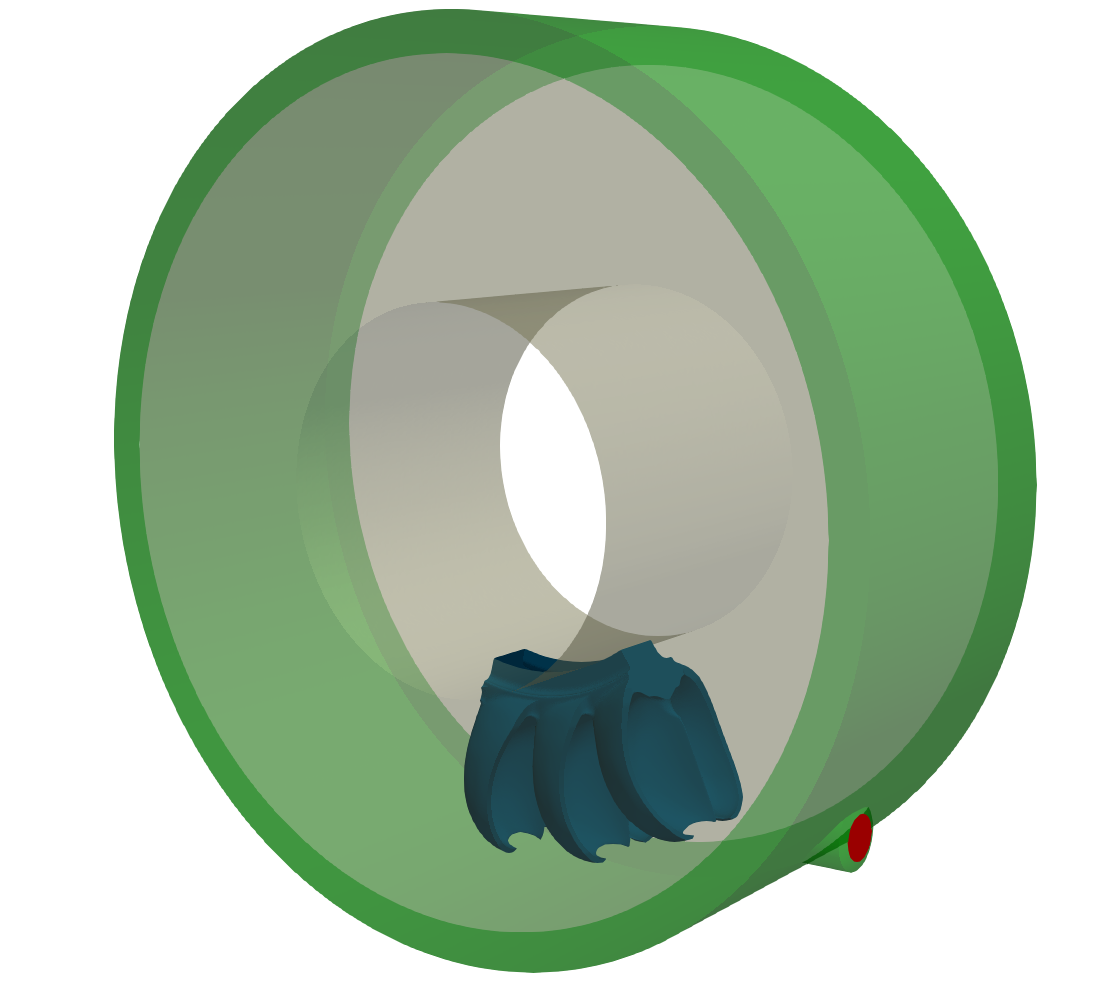}
        \caption{Pelton bucket domain, red zone is velocity inlet. The green indicates the static region and the light brown indicates the sliding mesh region, both are pressure outlets. The buckets are indicated in blue and are non-slip walls.}
    \end{subfigure}
    \hfill
    \begin{subfigure}[b]{0.4\textwidth}
    \centering
        \includegraphics[width=1\linewidth]{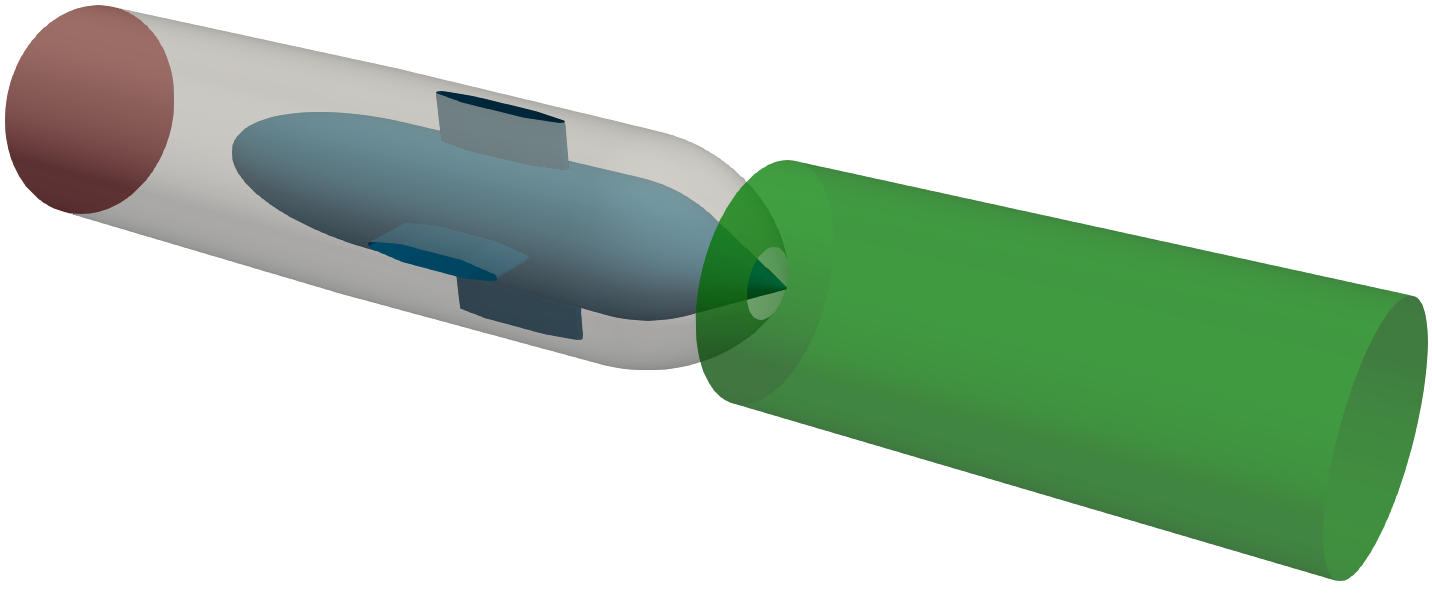}
        \caption{Nozzle simulation domain, red zone is inlet, white and blue zone are non slip walls and the green zone is outlet.}
    \end{subfigure}
    \caption{Boundary conditions}
    \label{fig:domain}
\end{wrapfigure}

\section{Methodology}
\subsection{Numerical scheme}
In this study, two simulation models are used, one of the jet impacting on the Pelton wheel and another to simulate the particle flow through the injector. For the study, the Eulerian-Lagrangian methodology is employed in Ansys Fluent \cite{ansys_fluent} where the water-air interface is solved with the volume of fluid (VOF) method and the sediment particles with the discrete phase model, (DPM) with one way coupling \cite{leguizamon_multiscale_2020}. Due to computational limitations, the Pelton wheel is simulated with only three buckets. The focus of the study is on the central bucket, while the side buckets are included to capture the interaction effects between adjacent buckets.
To analyse the effect of the Stokes number on the particle distribution in the nozzle, the DPM particles are solved with steady trajectories, while unsteady trajectories are solved in the simulation on the Pelton wheel. Turbulence is solved with RANS, k-$\omega$-SST with near wall correlation. The mass and momentum equations are discretized using a second-order upwind scheme and the volume fraction with the compressive scheme. For the simulation on the buckets, time advancement is conducted with an implicit scheme and an adaptive time-stepping is used in the simulation on the bucket with the Courant number of 3. For both simulations, a poly-hexcore mesh is employed with properties summarized in table \ref{tbl:mesh-quality}. The boundary conditions can be appreciated in figure \ref{fig:domain}. For the simulation of the Pelton wheel, gravity is parallel to the centre of rotation and for the simulation of the nozzle it is defined downwards.


\begin{table}
    \centering
    \caption{Mesh size and quality report}
    \label{tbl:mesh-quality}
    \begin{tabular}{c c c c c c c} \hline
        Simulation & No. Elements &  Min Ortho. & Avg Ortho. & Max Skewness & Avg Skewness & Avg. y$^+$ \\ \hline
        Pelton buckets & 12'970'346 & 0.281 & 0.972 & 0.719 & 0.0191 & 195\\ 
        Nozzle & 47'549'066 & 0.297 & 0.997 & 0.703 & 0.006 & 3.8\\ \hline
    \end{tabular}

\end{table}

\subsection{Sediment analysis}
A sediment analysis has been conducted based on a water sample taken at the Bitsch hydropower plant in Switzerland. The sediments composition is summarized in table \ref{tbl:mineralogy} and the size distribution of particles is illustrated in figure \ref{fig:part-dist}. As commonly accepted in literature to scale sediment laden flows ~\cite{li_effects_2018,solnordal_determination_2013}, we use the Stokes number (eq. \ref{eq:stokes}) to allow for comparison between different operational conditions, for various particle diameters as well as under different head and discharge conditions. We computed the Stokes number at the nozzle outlet so that it would be representative to the flow characteristics of interest in the nozzle itself. The Stokes number is defined as:

\begin{equation}
    Stk = \frac{\rho_p \cdot d_p^2\cdot u}{18 \cdot \mu \cdot D}
    \label{eq:stokes}
\end{equation}

where $\rho_p$ is the particle density, $d_p$, the particle diameter and $u$ the fluid velocity \cite{crowe_multiphase_2005}. In the equation, $\mu$ is the dynamic viscosity of the fluid and $D$ the outlet diameter of the nozzle.


\begin{wraptable}{r}{0.6\textwidth} 
    \centering
    \small
    \caption{Mineralogy results}
    \label{tbl:mineralogy}
    \begin{tabular}{c|c|c}
    Sediment type & Weight fraction [\%] & Density [kg/m$^3$] \\ \hline
    Quartz     & 53.14&  2650 \cite{noauthor_density_nodate} \\
    Phyllosilicates & 8.67 & \\
    Feldsparth  & 7.57  & 2590 \cite{noauthor_alkali_nodate}\\
    Plagioclase & 25.45 & 2680 \cite{noauthor_plagioclase_nodate}\\
    Calcite & 1.5 & 2710  \cite{noauthor_density_nodate-1} \\ 
    Dolomite & 1.29& 2780 \cite{noauthor_density_nodate}\\
    Undefined & 2.83 & \\ \hline
    Equivalent particle & 100 & 2646 \\ \hline
    \end{tabular}
\end{wraptable}

\section{Results}
\subsection{Particle distributions}
\begin{figure}[H]
    \centering
    \begin{subfigure}[b]{0.485\textwidth}
        \centering
        \includegraphics[width=1\linewidth]{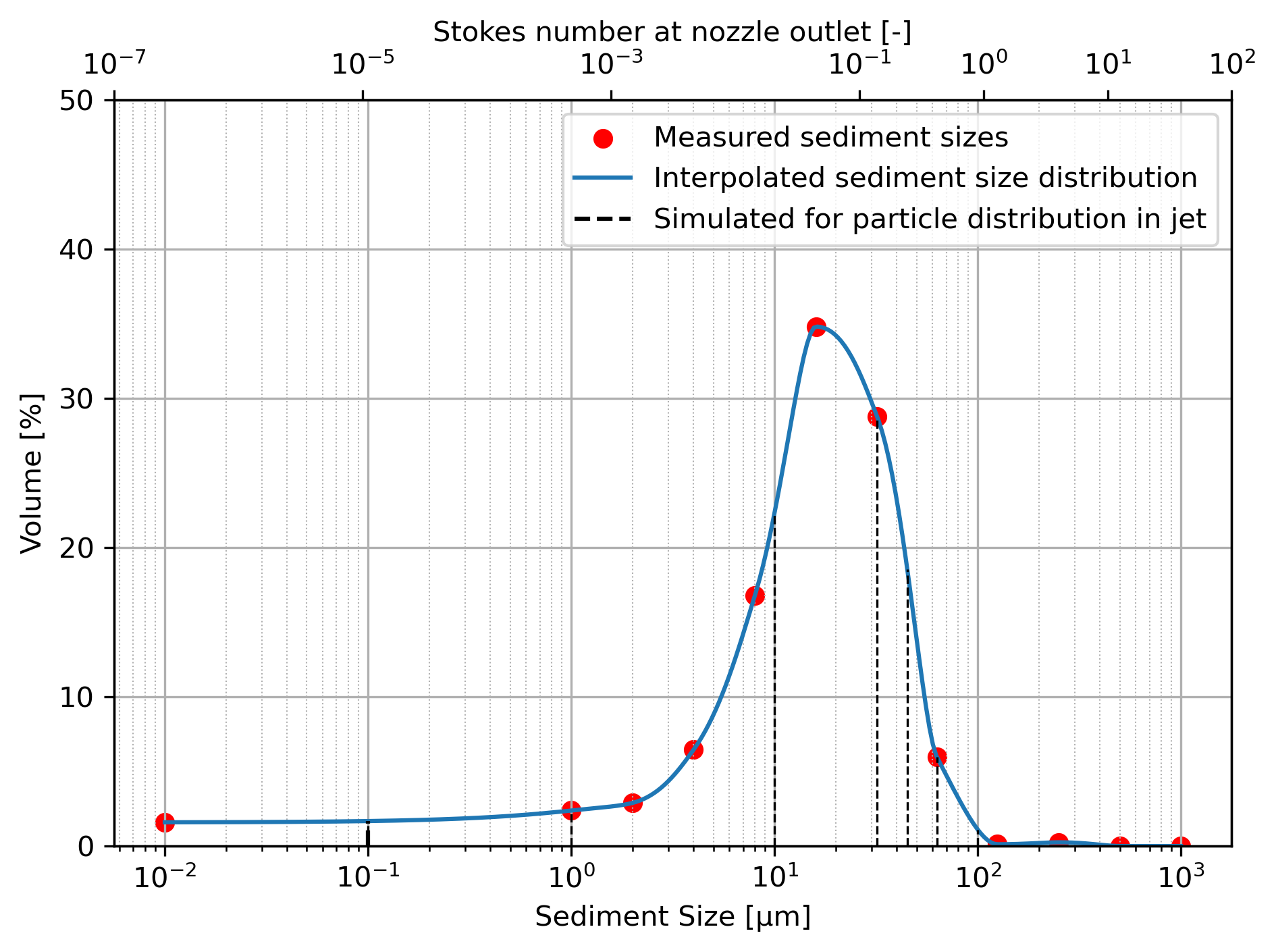}
        \caption{Particle size distribution in the sediment-laden flow running through Bitsch powerplant.}
        \label{fig:part-dist}
    \end{subfigure}
    \hfill
    \begin{subfigure}[b]{0.503\textwidth}
    \centering
        \includegraphics[width=1\linewidth]{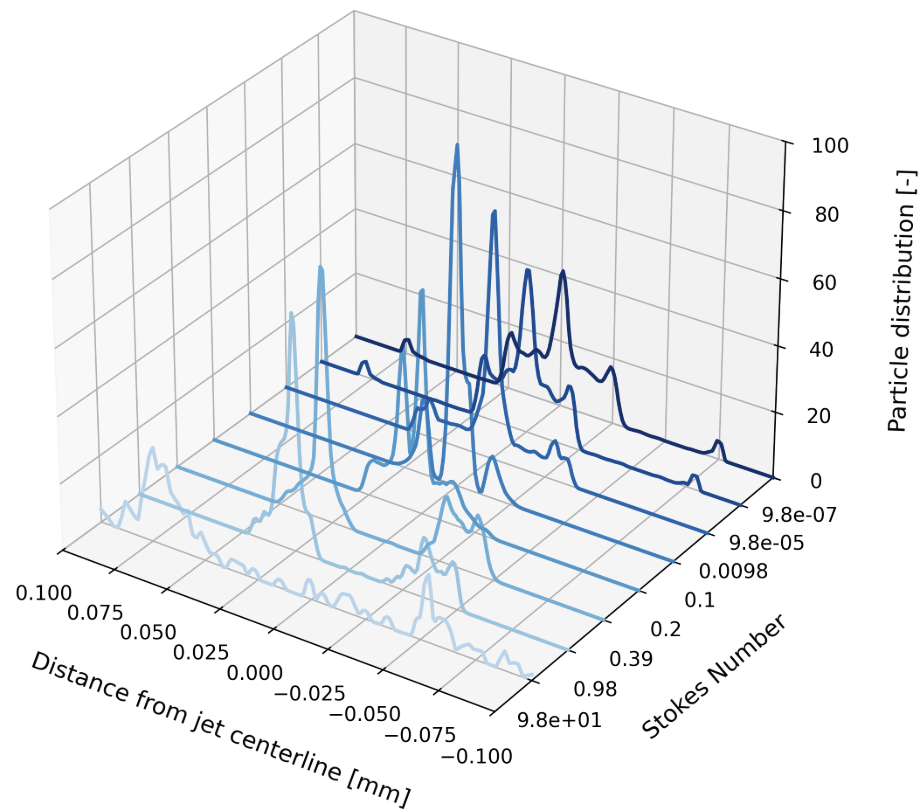}
        \caption{Sediment distribution along vertical line at jet centre, black line in figure \ref{fig:Stokes-nozzle} corresponds to sample location}
        \label{fig:jet-distribution}
    \end{subfigure}
    \label{fig:particles}
    \caption{Particle size and jet spatial distribution}
\end{figure}

The particle sizes and concentrations distribution in the sediment-laden flow running through Bitsch hydropower plant can be seen in figure \ref{fig:part-dist} together with the selected particle sizes for the numerical study. These sizes and corresponding concentrations are selected to investigate different Stokes numbers; 0.98, 0.39, 0.19, 0.1, 0.0098, 9.8e-05 and 5e-07. The effect of the Stokes number on the particle distribution in a cross-section of the jet at a distance of 4.2 the nozzle diameter can be seen in figure \ref{fig:jet-distribution}. For a Stokes number below $9.8 \times 10^{-5}$, it is observed that the particle distribution does not change significantly with further decreases in Stokes number. Additionally, for Stokes numbers below $0.10$, the particle distribution appears nearly axisymmetric around the jet axis. This axisymmetric distribution allows the particle concentration to be directly imposed as a function of the jet radius when simulating erosion on the Pelton buckets, thereby eliminating the need to simulate individual particle trajectories through the nozzle.
Since $93.7\,\%$ of the sediment particles in this study have a Stokes number below $0.1$, this simplification can be applied when imposing a realistic particle distribution in sediment-laden jet simulations. However, for higher Stokes numbers, the particle distribution within the jet is no longer axisymmetric, and particle velocities deviate from the jet direction. In such cases, it becomes necessary to couple the numerical simulations of the injector and the Pelton wheel to accurately define the spatial and velocity distribution of the particles.

\begin{figure}[H]
    \centering
    \includegraphics[width=1\linewidth]{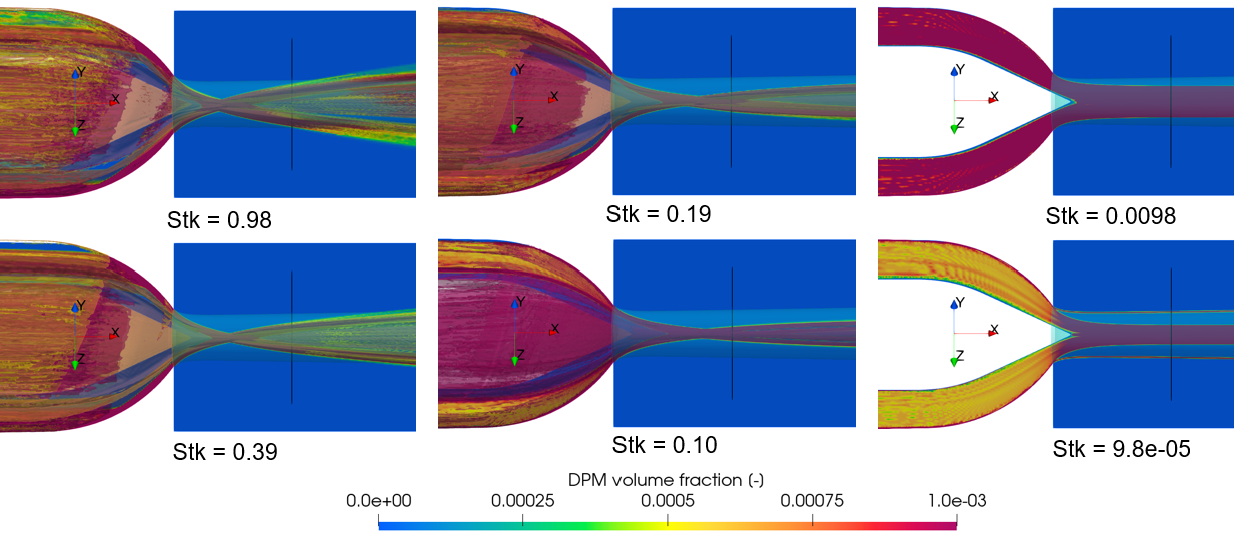}
    \caption{Effect of Stokes number on the particle trajectories, water-air interface of the jet is shown for reference in light blue. Black line indicates where particle distribution is taken for coupling with the simulation of the Pelton wheel.}
    \label{fig:Stokes-nozzle}
\end{figure}

The particle trajectories through the nozzle can be seen in figure \ref{fig:Stokes-nozzle}, where red means areas with high concentration of particles and blue areas where there are no particles. For high Stokes numbers such as those higher than 0.1, the asymmetry of the particle trajectories can be seen. For these sediment conditions, the influence of gravity plays an important role. These particles have high resistance to direction modification causing their trajectories to be defined based on collisions to the outer shell and the needle of the injector. It is also observed that at higher Stokes numbers, the particle trajectories initially converge toward a central point before diverging. This behaviour, caused by a higher ratio of momentum forces exerted on the particles compared to the drag forces, is of interest, as erosion damage correlates with both particle diameter and impact velocity. However, due to the divergence, some particles exit the jet and enter the surrounding low-velocity air region, resulting in deceleration before they impact the bucket.
At lower Stokes numbers (below $0.0098$), the particles follow the flow more closely and remain nearly parallel to it within the jet, resulting in a high concentration of particles near the jet centre. 

\subsection{Particles impact on Pelton bucket}
\begin{figure}[H]
    \centering
    \begin{subfigure}[b]{0.4\textwidth}
        \centering
        \includegraphics[width=1\linewidth]{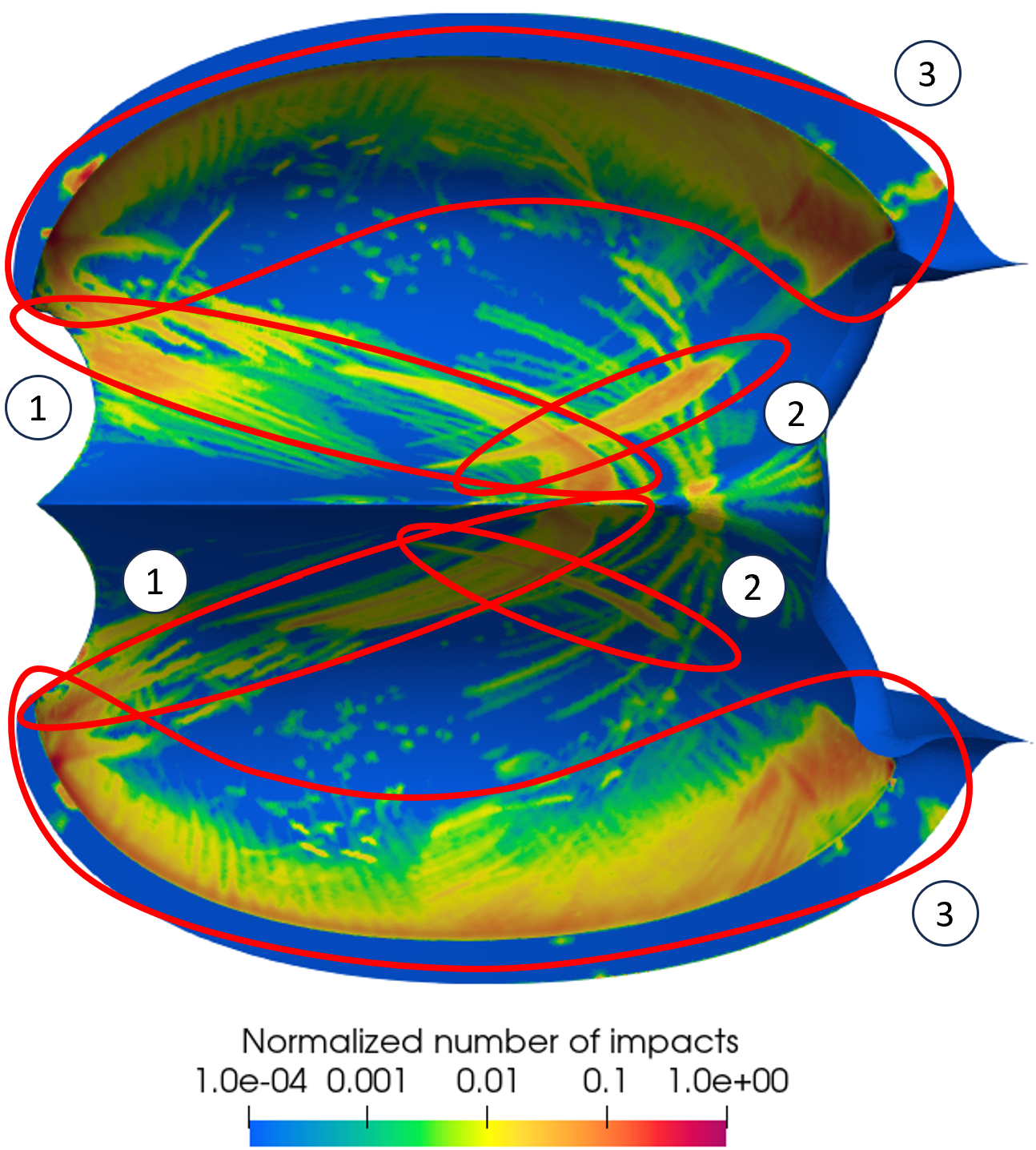}
        \caption{Ideal jet with uniform particle size and spatial distribution}
        \label{fig:bucket-ideal}
    \end{subfigure}
    \hfill
    \begin{subfigure}[b]{0.42\textwidth}
    \centering
        \includegraphics[width=1\linewidth]{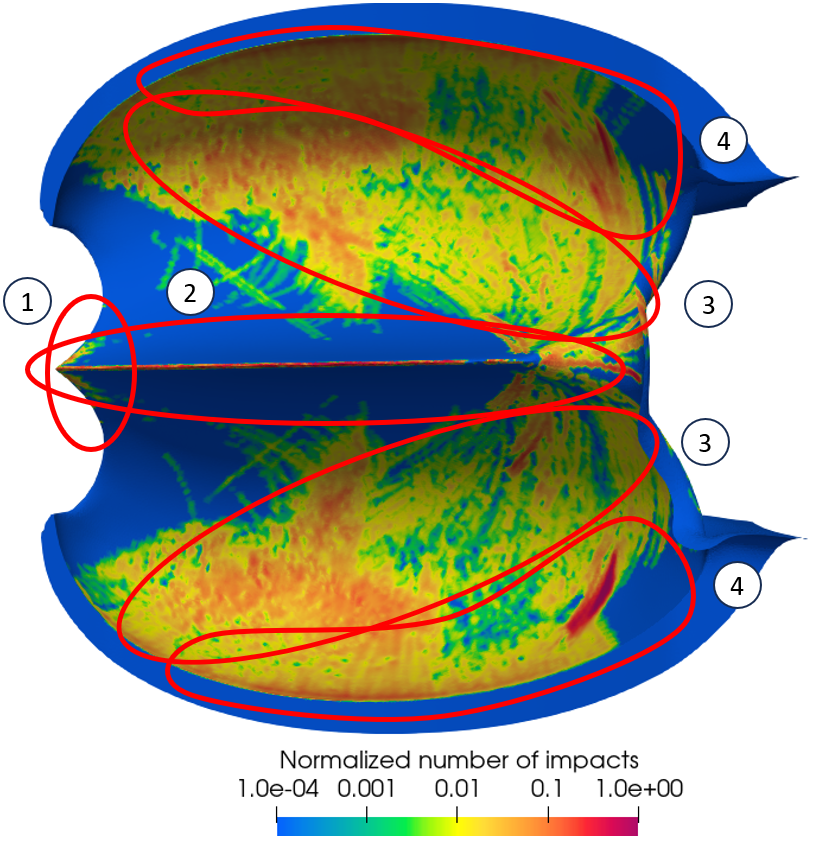}
        \caption{Corrected particle distribution with real size and spatial particle distribution in the jet}
        \label{fig:bucket-corrected}
    \end{subfigure}
    \caption{Comparison for normalized number of particle impacts for one jet passage between the ideal jet (left) and simulation results considering real particle distribution (right). }
    \label{fig:comparison-bucket}
\end{figure}

An injection file, which holds information on the particles in the flow i.e. the initial location and the velocity of them, was created for the inlet boundary condition of the simulation of the Pelton wheel to define the correct particle distributions in the jet corresponding to the Stokes numbers (fig. \ref{fig:part-dist}) present in the measurements taken at Bitsch powerplant. The file was created by dividing the sediment laden mass flow by the size distribution of sediment particles to get how many particles are for each particle size. For each particle size, the corresponding particle distribution was used to define the initial location of particles in the flow. Two simulations are conducted, one with an ideal jet with uniform particle distribution and an equivalent diameter and another with the corrected size and particle spatial distribution in the jet corresponding to the measured on-site sediments conditions. A comparison of the two considered cases is illustrated in figure \ref{fig:comparison-bucket}. For the ideal jet (fig. \ref{fig:bucket-ideal}), areas susceptible to erosion are marked by 1-3. Erosion zone 1 is caused by the jet entering the bucket whereas zones 2 and 3 are caused by the water leaving the bucket.  It is noticed that fewer number of impacts are observed in the bottom of the bucket compared to Ge et al. 2021 \cite{ge_experimental_2021} and  Leguizamón et al. 2020 \cite{leguizamon_multiscale_2020} which can be explained by their bucket Stokes number being two orders of magnitude larger than in our simulation. This is due to the fact that at lower Stokes numbers the particles follow the streamlines of the fluid more closely resulting in fewer impacts. For the corrected jet (fig. \ref{fig:bucket-corrected}), the areas susceptible to erosion are marked by 1-4. By imposing corrected particle distribution, an increase in impacts is noticed on the tip of the bucket (1) as well as on the splitter (2), areas where erosion is often under-predicted \cite{guo_analysis_2021} \cite{xiao_analysis_2022}. For the corrected particle distribution, more impacts were additionally observed in the bottom of the buckets (3) compared to the ideal jet. At the outlet of the bucket (4), the number of impacts has same magnitude in both simulations. The impact zones 3 and 4 are similar to the simulation of Xiao et al. 2022 \cite{xiao_analysis_2022}, where the nozzle flow was included in the simulation. However, the splitter and tip erosion (zones 1 and 2) where not observed in that article due to their assumption of a mean particle diameter for all particles.  

\section{Conclusion}

This study numerically demonstrates the influence of the Stokes number on particle distribution within the jet after it exits the injector. By using a more computationally efficient simulation methodology, coupling quick steady state injection simulations to demanding transient simulations of the buckets, it enhances the overall understanding of the erosion process on Pelton buckets by analysing how particle size and spatial distribution in the impacting jet affect the location and intensity of impact zones on the bucket surface. For low Stokes numbers ($<0.01$), the particle distribution follows a similar axisymmetric shape making it ideal to be imposed with a distribution function. This way, the probability of a particle being at a specific location in the jet can be imposed as a function of the jet radius. For high Stokes numbers ($>0.01$), gravity has a non-negligible influence making the particle distribution asymmetric. Due to this asymmetry, the particle distribution will have to be imposed by coupling simulations of the bucket with the injector. Furthermore, it was shown that by imposing the particle size and spatial distribution with a mean diameter and a uniform distribution will result in different erosion zones than when imposing the correct size and spatial distribution. This assumption was observed to cause under-prediction in the number of impacts on zones which are often heavily affected by erosion such as the tip of the bucket and the splitter. Future work will be done in validating this approach by comparing numerical models with experimental data for the same geometry. The authors acknowledge that solid particle erosion is not only characterised by the number of impacts but also the impact angle and velocity. Future works are foreseen also to include erosion models taking those parameters into account and compared to erosion data from field tests. This will help operators in including erosion forecasts into plant operation management and allowing for optimized sediments flushing through the turbine.

\section{Acknowledgment}
This work is framed within the ReHydro project and has received funding from the European Union’s Horizon Europe research and innovation programme under grant agreement No. 101147310 and COST Action PEN@Hydropower, CA21104, supported by COST (European Cooperation in
Science and Technology). The authors would like to thank the power-plant of Electra Massa as well as Martin Seydoux (Hydro Exploitation), Jean-Christophe Marongiu (Andritz), Martin Boden and Yann Le Cahain (Alpiq) for their valuable support in promptly providing on-site data and geometrical parameters for the case study.

\printbibliography[heading=bibintoc,title={References}]

\end{document}